\documentclass[12pt]{iopart}
\usepackage{graphicx}

\begin{document}

\title[Present status of controversies regarding the
thermal Casimir force]{Present status of controversies 
regarding the thermal Casimir force
}
\author{V~M~Mostepanenko$^{1}$,
V~B~Bezerra$^{2}$,
 R~S~Decca$^{3}$, B~Geyer$^{4}$,
E~Fischbach$^5$,
G~L~Klimchitskaya$^{6}$,
D~E~Krause$^{7,5}$,
D~L\'{o}pez$^{8}$ and C~Romero$^{2}$
}

\address{$^1$
Noncommercial Partnership  ``Scientific Instruments'', Moscow, Russia}
\address{$^2$
Department of Physics, Federal University of Paraiba, 
Jo\~{a}o Pessoa, Brazil}
\address{$^3$
Department of Physics, Indiana University-Purdue University
Indianapolis, Indianapolis, IN 46202, USA}
\address{$^4$
Center of Theoretical Studies and Institute for Theoretical Physics,
Leipzig University, Augustusplatz 10/11, D-04109 Leipzig, Germany 
}
\address{$^5$
Department of Physics, Purdue University,
West Lafayette, IN 47907, USA}
\address{$^6$
North-West Technical University, Millionnaya St. 5,
St.Petersburg, Russia
}
\address{$^7$
Department of Physics, Wabash College,
Crawfordsville, IN 47933, USA}
\address{$^8$ 
Bell Laboratories, Lucent Technologies,
Murray Hill, NJ 07974, USA}

\begin{abstract}
It is well known that, beginning in 2000, the behavior of
the thermal correction to the Casimir force between real metals
has been hotly debated. As was shown by several research groups,
the Lifshitz theory, which provides the theoretical foundation
for the calculation of both the van der Waals and Casimir
forces, leads to different results depending on the model
of metal conductivity used. To resolve these controversies,
the theoretical considerations based on the principles
of thermodynamics and new experimental tests were invoked.
We analyze the present status of the problem (in particular,
the advantages and disadvantages of the approaches based on
the surface impedance and on the Drude model dielectric function)
using rigorous analytical calculations of the entropy of
a fluctuating field.
We also discuss the results of a new precise experiment
on the determination of the Casimir pressure between two
parallel plates by means of a micromechanical torsional
oscillator. 
\end{abstract}
\pacs{12.20.Fv, 12.20.Ds, 11.10.Wx, 05.70.-a}
%\submitto{\JPA}
%\maketitle

\section{Introduction: Lifshitz formula provides the theoretical
background for the thermal Casimir force}

Since the
beginning of 2000, the dependence of the Casimir force acting
between real metals on separation and temperature has been hotly
debated. It is well known that van der Waals and Casimir
forces act between closely-spaced macroscopic bodies. Both
forces are of the same physical nature and are caused by 
long wavelength electromagnetic fluctuations \cite{1}.
The name ``van der Waals'' is usually used in a nonrelativistic
case when the force and other physical quantities depend on
the Planck constant $\hbar$, but do not depend on the velocity
of light $c$. In the relativistic case, which corresponds to
larger separations, the name ``Casimir'' is commonly used, 
since he was the first to predict the force acting in vacuum
between two parallel uncharged plates made of ideal metals
\cite{2}. During the last few years dispersion forces (which
is the generic name for both kinds of forces caused by
fluctuations) were actively investigated, not only
theoretically but also experimentally (see the monographs
[3--5] and a review \cite{6}). This was motivated by many
prospective applications of dispersion phenomena in
fundamental physics and nanotechnology. In particular, the
problem of an accurate theoretical description of dispersion
forces between real bodies (i.e., taking into account the
realistic conductivity properties, surface roughness and nonzero
temperature) took on great significance. This paper is
devoted to the outstanding problems arising in the case of
plates made of real metals and having nonzero temperature.

The free energy of the dispersion interaction between two thick
parallel plates (semispaces) described by a dielectric
permittivity $\varepsilon(\omega)$, with a gap of width $z$
between them in thermal equilibrium at temperature $T$, was
derived by Lifshitz \cite{7}. It can be represented in the
form
\begin{eqnarray}
&&
{\cal F}(z,T)=\frac{k_BT}{2\pi}
\sum\limits_{l=0}^{\infty}
\left(1-\frac{1}{2}\delta_{l0}\right)
\int_{0}^{\infty}k_{\bot}dk_{\bot}
\label{eq1} \\
&&
\phantom{aaa}\times
\left\{\ln\left[1-r_{\|}^2(\xi_l,k_{\bot})e^{-2q_lz}\right]+
\ln\left[1-r_{\bot}^2(\xi_l,k_{\bot})e^{-2q_lz}\right] 
\right\}.
\nonumber
\end{eqnarray}
\noindent
Here $k_B$ is the Boltzmann constant,
$\xi_l=2\pi k_BTl/\hbar$ are the Matsubara frequencies,
$q_l=(k_{\bot}^2+\xi_l^2/c^2)^{1/2}$, $k_{\bot}$ is
the projection of the wave vector in the plane of the plates,
and $r_{\|,\bot}(\xi_l,k_{\bot})$ are the reflection 
coefficients for two independent polarizations of the 
electromagnetic field. The explicit form of these
coefficients in the case of real metals is closely
connected to the controversies which are the
subject of this paper.

\section{Two main approaches to the presentation of
reflection coefficients: the dielectric permittivity
and the surface impedance}

In the framework of the first approach, the reflection
coefficients are expressed in terms of the dielectric
permittivity of a plate material. In application to
Eq.~(\ref{eq1}) this approach goes back to 
E.{\ }M.{\ }Lifshitz who used it to describe the
Casimir and van der Waals forces between dielectrics.
In this case
\begin{equation}
r_{\|}\equiv r_{\|}^L(\xi_l,k_{\bot})=
\frac{\varepsilon_lq_l-k_l}{\varepsilon_lq_l+k_l},
\quad
r_{\bot}\equiv r_{\bot}^L(\xi_l,k_{\bot})=
\frac{k_l-q_l}{k_l+q_l},
\label{eq2}
\end{equation}
\noindent
where $k_l=(k_{\bot}^2+\varepsilon_l\xi_l^2/c^2)^{1/2}$
and $\varepsilon_l\equiv\varepsilon(i\xi_l)$.
Within the Lifshitz approach, the fluctuating
electromagnetic field is considered both in the gap and
in the interior of the dielectric semispaces with the
continuity boundary conditions on their surfaces
\begin{equation}
E_{1t}=E_{2t}, \quad
B_{1t}=B_{2t}, \quad
D_{1n}=D_{2n}, \quad
B_{1n}=B_{2n}
\label{eq3}
\end{equation}
\noindent
(note that we do not consider ferromagnets and set
{\boldmath$B$}$=${\boldmath$H$}).

Of prime importance is the contribution of the free energy in 
Eq.~(\ref{eq1}) with $l=0$. It is well known that this
term determines the total value of ${\cal F}$ at high
temperatures (large separations) because all
contributions with $l\geq 1$ are exponentially small
in this limiting case \cite{5,6}. Eq.~(\ref{eq1}) with
reflection coefficients (\ref{eq2}) correctly describes 
the case of plates made of ideal metals if one
uses the so-called Schwinger prescription \cite{5,6},
i.e., take limit $\varepsilon\to\infty$ first and set
$l=0$ afterwards. In doing so for ideal metals
the equations
\begin{equation}
r_{\|}^L(0,k_{\bot})=r_{\bot}^L(0,k_{\bot})=1
\label{eq4}
\end{equation}
\noindent
are obtained which coincide with those obtained
 from thermal
quantum field theory with Dirichlet boundary
conditions \cite{8} independent of the Lifshitz formula.

In Ref.~\cite{9} an attempt was undertaken to describe
the dispersion interaction between real metals by the
Lifshitz formula using the Drude model to
characterize their dielectric properties at low
frequencies:
\begin{equation}
\varepsilon(i\xi)=
1+\frac{\omega_p^2}{\xi\left[\xi+\gamma(T)\right]}
\approx 1+\frac{\omega_p^2}{\xi\gamma(T)},
\label{eq5}
\end{equation}
\noindent
where $\omega_p$ is the plasma frequency and $\gamma(T)$ is
the relaxation parameter. Substituting Eq.~(\ref{eq5}) into
Eq.~(\ref{eq2}) one obtains
\begin{equation}
r_{\|}^L(0,k_{\bot})=1,
\quad
r_{\bot}^L(0,k_{\bot})=0.
\label{eq6}
\end{equation}
\noindent
Eq.~(\ref{eq6}) is preserved also in the limit of ideal
metals, and is thus in contradiction with Eq.~(\ref{eq4}).
If, however, one characterizes a metal by the plasma
model,
\begin{equation}
\varepsilon(i\xi)=1+\frac{\omega_p^2}{\xi^2},
\label{eq7}
\end{equation}
\noindent
and extrapolates it to low frequencies, 
the reflection coefficients become:
\begin{equation}
r_{\|}^L(0,k_{\bot})=1,
\quad
r_{\bot}^L(0,k_{\bot})=
\frac{\sqrt{c^2k_{\bot}^2+\omega_p^2}-
ck_{\bot}}{\sqrt{c^2k_{\bot}^2+\omega_p^2}+ck_{\bot}}.
\label{eq8}
\end{equation}
\noindent
Eq.~(\ref{eq8}) agrees with the case of ideal
metals (\ref{eq4}) because $r_{\bot}^L(0,k_{\bot})\to 1$
when $\omega_p\to\infty$.

The approach to the theoretical description of the thermal
Casimir force between real metals based on Eq.~(\ref{eq6})
was supported and developed in Refs.~[10--13], while that
following Eqs.~(\ref{eq7}) and (\ref{eq8}) was proposed in
Refs.~\cite{15,16}. By convention we refer to them below
as ``the Drude model approach'' and ``the plasma model
approach'', respectively.

The Drude model approach predicts relatively large 
thermal corrections to the Casimir force between real metals 
at short separations of about 500 times greater than between 
ideal metals. At large separations or high temperatures the
Drude model approach predicts 1/2 the magnitude of
the Casimir free energy and force than between ideal metals. 
(The same prediction was obtained using the model of a
conducting wall by the classical Coulomb fluid \cite{15a}
or ``non-relativistic quantum electrodynamics''\cite{15b}.)
This prediction is in contradiction with the classical limit
based on Kirchhoff's law \cite{15c}. As to the
plasma model approach, its predictions are in qualitative
agreement with the case of ideal metals. Because of this,
the plasma model approach is called ``traditional'' and
the Drude model approach is called ``alternative''. The
apparent advantage of the Drude model approach is that
Eq.~(\ref{eq5}) presents the same behavior of
$\varepsilon\sim\omega^{-1}$ at low frequencies,
as given by the Maxwell equations, whereas the behavior
$\varepsilon\sim\omega^{-2}$ in Eq.~(\ref{eq7}) is
characteristic for the region of infrared optics, raising
a question on the correctness of its extrapolation to
low frequencies. On the other hand, Eq.~(\ref{eq8}) is
in agreement with the limiting case of an ideal metal,
whereas Eq.~(\ref{eq6}) is not. Bearing in mind that
the description of real metals using
$\varepsilon$ depending only on $\omega$ is not
universal, it is worthwhile to use alternative
physical quantities to characterize the reflection 
properties of metal surfaces.

In the framework of the second approach, which goes
back to V.{\ }L.{\ }Ginzburg and his collaborators
(see, e.g., Ref.~\cite{17}), the reflection coefficients
in the Lifshitz formula are expressed in terms of
the Leontovich impedance $Z(\omega)$. The Leontovich
impedance boundary condition
\begin{equation}
{\mbox{\boldmath$E$}}_t=Z(\omega)\left[
{\mbox{\boldmath$H$}}_t\times{\mbox{\boldmath$n$}}\right]
\label{eq9}
\end{equation}
\noindent
(where {\boldmath$n$} is an internal unit vector 
directed into the medium and normal to the boundary) is 
approximate and permits one to find the electromagnetic
field in the gap without considering it in the interior
of the plates. In terms of the Leontovich impedance,
the reflection coefficients in Eq.~(\ref{eq1}) are
given by
\begin{equation}
r_{\|}\equiv r_{\|}^G(\xi_l,k_{\bot})=
\frac{cq_l-Z_l\xi_l}{cq_l+Z_l\xi_l},
\quad
r_{\bot}\equiv r_{\bot}^G(\xi_l,k_{\bot})=
\frac{\xi_l-cq_lZ_l}{\xi_l+cq_lZ_l},
\label{eq10}
\end{equation}
\noindent
where  $Z_l\equiv Z(i\xi_l)$. The Leontovich impedance
is applicable even in some cases when
$\varepsilon(\omega)$ loses its meaning (e.g., in the frequency
region of the anomalous skin effect). If, however, both
quantities are in their domain of application, they are
related by $Z(\omega)=1/\sqrt{\varepsilon(\omega)}$
\cite{18}. For the impedance functions of the normal and
anomalous skin effect, the Ginzburg reflection coefficients at
zero frequency satisfy Eq.~(\ref{eq4}) as do the Lifshitz ones
for ideal metals.
In the region of infrared optics it follows that 
\begin{equation}
r_{\|}^G(0,k_{\bot})=1,
\quad
r_{\bot}^G(0,k_{\bot})=
\frac{\omega_p-ck_{\bot}}{\omega_p+ck_{\bot}}.
\label{eq11}
\end{equation}
\noindent
The impedance approach to the thermal Casimir force was
developed in Refs.~[21--24]. It predicts small thermal
corrections at short separations in agreement with the
case of ideal metals and leads to the same high-temperature
(large separation) asymptotic results which hold for the
plasma model approach or for ideal metals in accordance
with the classical limit. Thus, the
impedance approach can also be called ``traditional''.

Many computations of the thermal Casimir force have been
performed in the framework of the Drude model approach
[10--13], plasma model approach \cite{15,16} and impedance
approach [21--24], and many controversial statements
in favor and against each approach have appeared
in the literature. These
arguments were most recently presented in Refs.~\cite{23,24}.
Before discussing them in more detail, we emphasize that
the Drude model and the impedance approaches are qualitatively
different only at zero Matsubara frequency [compare
Eqs.~(\ref{eq6}) and (\ref{eq11})]. As to the contributions
from Matsubara frequencies with $l\geq 1$, both approaches
find these using tabulated optical data extrapolated
to low frequencies by the imaginary part of the Drude
dielectric function with practically coinciding results.
The plasma model approach applied at all frequencies
disregards the internal photoelectric effect (interband
transitions) and other processes taken into account
in the optical data. Because of this, it is not as accurate
as the impedance approach at short separations of
about 150--250\,nm.

\section{Controversies between the two approaches}

\subsection{Is there a violation of thermodynamics for the
Lifshitz formula combined with the Drude model?}

As was proven analytically in Ref.~\cite{22}, the Drude
model approach leads to a violation of the third law of
thermodynamics (the Nernst heat theorem) in the case of
metallic perfect lattices with no defects or impurities.
For such lattices the relaxation parameter $\gamma(T)\to 0$
when $T\to 0$ in accordance with the Bloch-Gr\"{u}neisen
law (this property is preserved even when the effects of
electron-electron collisions are included), and the
entropy of a fluctuating field takes the form
\begin{equation}
\fl
S(z,0)=-\frac{\partial{\cal F}(z,T)}{\partial T}{\vert}_{T=0}
=\frac{k_B}{16\pi z^2}\int_{0}^{\infty}{\!\!\!}ydy
\ln\left[1-\left(\frac{y-\sqrt{\frac{\omega_p^2}{\omega_c^2}+y^2}}{y+
\sqrt{\frac{\omega_p^2}{\omega_c^2}+y^2}}\right)^2e^{-y}\right]<0,
\label{eq12}
\end{equation}
\noindent
where $\omega_c=c/(2z)$ is the characteristic frequency of the
Casimir effect.

To avoid this conclusion, Refs.~\cite{12,14} apply the Drude
model approach to metallic lattices with defects and
impurities possessing some residual relaxation $\gamma(0)\neq 0$
and obtain $S(z,0)=0$. This, however, does not solve the
problem of the thermodynamic inconsistency of the Drude model
approach, because perfect lattices have a nondegenerate dynamical
state of lowest energy and, thus, according to quantum statistical
physics, the entropy at zero temperature must be
equal to zero in this case
[a property violated by the Drude model approach
according to Eq.~(\ref{eq12})]. Recently
the authors of Ref.~\cite{24} recognized
that the Drude model approach violates thermodynamics ``for
perfect metals of infinite extension'' having no relaxation
(the so-called ``modified ideal metal'' which is obtained from
the real metals in the Drude model approach when 
$\varepsilon\to\infty$ \cite{24}), but they deny a 
violation for metals which include relaxation. This denial
is based on a misunderstanding. As was already mentioned 
above, the Drude
model approach violates the Nernst heat theorem for metallic
perfect lattices with no defects or impurities [see
Ref.~\cite{22} and Eq.~(\ref{eq12})]. Such lattices have
a nonzero resistance and relaxation at any nonzero temperature,
which go to zero only when temperature vanishes. Thus, they
are not ``perfect metals'' according to Ref.~\cite{24}.

In the light of the above, the conclusion of Refs.~\cite{12,24} 
that the Drude model approach is in agreement with thermodynamics
cannot be supported. In contrast, the plasma model and the
impedance approaches are in complete agreement with
thermodynamics \cite{16,22}.

\subsection{Do the ``exact'' impedances lead to the same result
as the Drude model or as the Leontovich impedance?}

As was mentioned in Sec.~2, the impedance boundary condition
(\ref{eq9}) is an approximate one. In particular, it works well
if the electromagnetic waves penetrating into metal interior
propagate almost perpendicular to its surface. (This is in fact
the case as long as $|\varepsilon(\omega)|\gg 1$.)
Some authors (see, e.g., Refs.~\cite{12,25}) introduce 
so-called ``exact'' impedances
\begin{equation}
\fl
Z_{\|}(\omega,k_{\bot})=Z(\omega)
\left(1-\frac{c^2k_{\bot}^2}{\omega^2\varepsilon}\right)^{1/2},
\quad
Z_{\bot}(\omega,k_{\bot})=Z(\omega)
\left(1-\frac{c^2k_{\bot}^2}{\omega^2\varepsilon}\right)^{-1/2}, 
\label{eq12a}
\end{equation}
\noindent
which depend on both polarization and transverse momentum,
and related impedance boundary conditions
\begin{equation}
{\mbox{\boldmath$E$}}_t=Z_{\bot}(\omega,k_{\bot})\left[
{\mbox{\boldmath$H$}}_t\times{\mbox{\boldmath$n$}}\right],
\quad
Z_{\|}(\omega,k_{\bot}){\mbox{\boldmath$H$}}_t=\left[
{\mbox{\boldmath$n$}}\times{\mbox{\boldmath$E$}}_t\right],
\label{eq13}
\end{equation}
\noindent
where $Z(\omega)$ is the Leontovich impedance.
Actually, the impedances (\ref{eq12a}) are not exact.
Eqs.~(\ref{eq12a}) and (\ref{eq13}) follow from the Maxwell
equations only if $\varepsilon(\omega)$ has definite meaning,
so that the spatial dispersion is neglected and the relation
{\boldmath$D$}({\boldmath$r$},$\omega$)$=
\varepsilon(\omega)${\boldmath$E$}({\boldmath$r$},$\omega$)
is valid. If, however, the homogeneity of space is
violated by the presence of boundaries, 
and if  spatial dispersion is taken into 
account,  
Eqs.~(\ref{eq12a}) and (\ref{eq13}) are in general inapplicable
(see Sec.~5).

In terms of the impedances (\ref{eq12a}), the reflection
coefficients take the form
\begin{equation}
r_{\|}(\xi_l,k_{\bot})=
\frac{cq_l-Z_{l,\|}\xi_l}{cq_l+Z_{l,\|}\xi_l},
\quad
r_{\bot}(\xi_l,k_{\bot})=
\frac{\xi_l-cq_lZ_{l,\bot}}{\xi_l+cq_lZ_{l,\bot}},
\label{eq14}
\end{equation}
\noindent
where $Z_{l,\|}\equiv Z_{\|}(i\xi_l,k_{\bot})$ and
$Z_{l,\bot}\equiv Z_{\bot}(i\xi_l,k_{\bot})$.
It is easy to check that if one substitutes Eq.~(\ref{eq12a})
into Eq.~(\ref{eq14}), considers $\xi_l$ and $k_{\bot}$
as independent variables, and takes into account that
$Z=1/\sqrt{\varepsilon}$, the Lifshitz reflection coefficients
(\ref{eq2}) are recovered. This allowed the authors
of Refs.~\cite{11,12,24} to insist that the results (\ref{eq4})
and (\ref{eq11}), obtained in Refs.~\cite{20,22} within the
impedance approach [in contrast to Eq.~(\ref{eq6}) valid in
the Drude model approach], are due to disregarding the transverse
momentum dependence by the Leontovich impedance.

Bearing in mind that Eq.~(\ref{eq6}) returns us to the 
contradiction with thermodynamics, it is necessary to analyze
this point more thoroughly. When we are considering 
real photons incident on a plane boundary of a single
semispace, the components of a wave vector are constrained
by the mass-shell equation
$|{\mbox{\boldmath$k$}}|^2=k_{\bot}^2+k_3^2=\omega^2/c^2$
($k_3$ is the wave vector component perpendicular to the
boundary). Using this equation, the angle of incidence
is expressed by 
$\sin\vartheta_0=k_{\bot}/|\mbox{\boldmath$k$}|=
ck_{\bot}/\omega$ and
the impedances (\ref{eq12a}) can be identically rewritten
in the form
\begin{equation}
\fl
Z_{\|}(\omega,k_{\bot})=Z(\omega)
\left[1-\frac{\sin^2\vartheta_0}{\varepsilon(\omega)}\right]^{1/2},
\quad
Z_{\bot}(\omega,k_{\bot})=Z(\omega)
\left[1-\frac{\sin^2\vartheta_0}{\varepsilon(\omega)}\right]^{-1/2}. 
\label{eq15}
\end{equation}
\noindent
The term ${\sin^2\vartheta_0}/{\varepsilon(\omega)}$ can be
neglected for all frequencies which are at least several times
smaller than the plasma frequency because in this region 
$|\varepsilon|\gg 1$. Furthermore, in the limit of zero frequency
${\sin^2\vartheta_0}/{\varepsilon}\to 0$ due to 
$|\varepsilon|\to\infty$, and as a result for real photons
the Leontovich
impedance coincides with the impedances $Z_{\|,\bot}$ precisely.
The mass-shell equation ensures the fulfilment of the Snell's
laws. Thus, if the reflection properties for virtual photons
are postulated to be the same as for real photons, the Leontovich
impedance and boundary condition (\ref{eq9}) are recovered.

These considerations can be generalized for the case of two
semispaces with the boundary conditions (\ref{eq13}).
In this case the Maxwell equations lead to the
following dispersion relations for the determination of
photon eigen-frequencies $\omega$ \cite{20},
\begin{eqnarray}
&&
\Delta_{\|}(\omega,k_{\bot})\equiv e^{-qz}\left(1-\eta_{\|}^2\right)
\left(\sinh qz-\frac{2i\eta_{\|}}{1-\eta_{\|}^2}
\cosh qz\right)=0,
 \label{eq16} \\
&&
\Delta_{\bot}(\omega,k_{\bot})\equiv e^{-qz}
\left(1-\kappa_{\bot}^2\right)
\left(\sinh qz+\frac{2i\kappa_{\bot}}{1-\kappa_{\bot}^2}
\cosh qz\right)=0,
\nonumber
\end{eqnarray}
\noindent
where
\begin{equation}
q^2=k_{\bot}^2-\omega^2/c^2,
\quad
\eta_{\|}=\frac{\omega Z_{\|}(\omega,k_{\bot})}{cq},
\quad
\kappa_{\bot}=\frac{cqZ_{\bot}(\omega,k_{\bot})}{\omega}.
\end{equation}
\noindent
It is easily seen from Eq.~(\ref{eq16}) that in the limit
$\omega\to 0$ the transverse momentum 
must also go to zero, $k_{\bot}\to 0$, in such a way that
$ck_{\bot}\sim\omega$. This means that the
quantity $c^2k_{\bot}^2/(\omega^2\varepsilon)$ in 
Eq.~(\ref{eq12a}) goes to zero in the limit of zero frequency,
and both impedances coincide precisely with the Leontovich
impedance. At all nonzero frequencies there
are only negligible differences between the impedances 
(\ref{eq12a}) and the Leontovich impedance. Thus, in the
case of two semispaces, the dispersion relations (\ref{eq16})
play the same role as the mass-shell equation for free
photons.

Using the above property of the solutions of Eq.~(\ref{eq16}),
this equation can be approximately rearranged to the form 
following from the boundary conditions (\ref{eq9}),
\begin{eqnarray}
&&
\Delta_{\|}(\omega,k_{\bot})\equiv e^{-qz}\left(1-\eta^2\right)
\left(\sinh qz-\frac{2i\eta}{1-\eta^2}
\cosh qz\right)=0,
 \label{eq18} \\
&&
\Delta_{\bot}(\omega,k_{\bot})\equiv e^{-qz}
\left(1-\kappa^2\right)
\left(\sinh qz+\frac{2i\kappa}{1-\kappa^2}
\cosh qz\right)=0,
\nonumber
\end{eqnarray}
\noindent
where $\eta=\omega Z/(cq)$, $\kappa=cqZ/\omega$ are now
expressed in terms of the Leontovich impedance. Note that the
approximation was made only at $\omega\neq 0$, whereas
in the limit $\omega\to 0$, which is of special importance for our
problem, Eqs.~(\ref{eq16}) and (\ref{eq18}) remain equivalent.

It is now easy to proceed with the derivation of the
Lifshitz formula by summing the free energies of all
oscillators whose frequencies are determined by equations
(\ref{eq18}), applying the argument theorem and calculating 
the residues at imaginary Matsubara frequencies \cite{20}. 
This derivation results in Eq.~(\ref{eq1}) with the
reflection coefficients in Eq.~(\ref{eq10}). Note that in our
derivation the transverse momentum dependence is not
``completely disregarded'' (as is claimed in 
Refs.~\cite{11,12,24}). The transverse momentum $k_{\bot}$
enters through the quantity $q$ both in the dispersion
equations (\ref{eq18}) and in the reflection coefficients
(\ref{eq10}). We emphasize that the dispersion relations
(\ref{eq18}) with the Leontovich impedance have both photonic 
and plasmonic solutions. It is not admissible, however,
to preserve $k_{\bot}\neq 0$ in the impedance function
$Z_{\bot}(\omega,k_{\bot})$, and to consider $\omega\to 0$
because this violates the dispersion relations (\ref{eq16}).
The Matsubara frequencies and the transverse momentum
become independent only after application of the
argument theorem.
In view of the above, the conclusion of
Ref.~\cite{12} that the Drude model and impedance
approaches lead to the same reflection coefficients at
zero frequency is not warranted.

\subsection{How to extrapolate real data to zero frequency?}

As was already discussed in Sec.~2, the tabulated optical data
for real materials do not contain the zero frequency
 contribution which should be found from some theoretical
considerations. There are two main methods proposed in
literature for solving this problem. According to one
method (see, e.g., Refs.~\cite{12,24}) it is necessary to
extrapolate to zero the actual reflection properties of plate 
materials at very low, quasistatic frequencies.
According to another method, the zero-frequency limit
should not be understood literally because there are no
static-field fluctuations. More likely it should be
understood as a mathematical limit to zero from the region
around the characteristic frequency $\omega_c$ giving the
major contribution to the thermal Casimir force [21--24].
Although in Ref.~\cite{12} such an extrapolation is considered
to be ``ad hoc'', in Ref.~\cite{21} it was demonstrated
that the method of Refs.~\cite{12,24}, applied in the framework 
of the impedance approach, results in a violation of the 
Nernst heat theorem.

Recently the method of Refs.~[21--24] received support when
applied to two dielectric semispaces,
i.e., to the case which was previously free of any
controversies. As was analytically proved in Ref.~\cite{26}
for similar (and in Ref.~\cite{27} for dissimilar) dielectrics,
the Casimir entropy of a fluctuating field goes to zero when 
temperature vanishes if the semispace materials are
described by a frequency dependent $\varepsilon$ with
some finite static values $\varepsilon(0)$. If, however,
the dc conductivity of dielectrics is taken into
account, this results in a violation of the Nernst heat 
theorem. When it is recalled that dielectrics really possess
some nonzero conductivity at a constant current (although
it is many orders of magnitudes smaller than that for metals),
it becomes clear that real material properties at very low
frequencies are in fact irrelevant to the fluctuating
phenomena and must not be included into the model of the
dielectric response.

\section{Different approaches to the thermal Casimir force
and experiment}

\subsection{Measurement of the Casimir force using a
torsion pendulum}

As was mentioned in Sec.~2, the traditional approaches to
the thermal Casimir force predict small thermal corrections
at short separations in qualitative agreement with the case
of ideal metals. The sensitivity of all already performed
experiments is not sufficient to measure such corrections,
much less to discriminate among different traditional
approaches. Some of these experiments are, however, 
sufficiently sensitive to detect the alternative
thermal corrections as predicted by the Drude model approach
if they exist.

According to the results of the first modern measurement of
the Casimir force between a Au coated plate and large
spherical lens \cite{28}, at $T=300\,$K with $z=1\,\mu$m, a 
net deviation between the Casimir forces computed for ideal
metals and using the Drude model approach is 25\%. Of this
deviation, 19\% is due to the large alternative thermal
correction. In Ref.~\cite{28} the experimental uncertainty
at $z=1\,\mu$m was about 5--10\%. The predicted thermal
effect was not, however, observed. According to Ref.~\cite{29},
the prediction of the Drude model approach ``deviates
significantly from experimental results''. At the same
time, according to Ref.~\cite{30}, the experimental data are
consistent with the impedance approach.

\subsection{Static experiment using a microelectromechanical
torsional oscillator}

In this experiment the Casimir force between a plate of a
torsional oscillator and a sphere was measured \cite{31}. The
sphere was coated by Au and the plate by Cu. The force values
were measured with an absolute error of 0.3\,pN for
separations in the range 0.19--1.2$\,\mu$m. Ref.~\cite{12}
claims that there is ``reasonably good agreement'' between
the Drude model approach and the experimental results of the
static experiment at $z=200\,$nm.
In Ref.~\cite{23} it was demonstrated, however, that this
conclusion is based on a misunderstanding of the data. 
In fact, the
experimental data are in contradiction with the Drude model 
approach, but are consistent with the traditional approaches to
the thermal Casimir force. For example, at about
$z=200\,$nm the mean value of $F^{\rm th}-F^{\rm exp}$
(where the theoretical Casimir force $F^{\rm th}$ is
computed using one of the traditional approaches) is
practically zero. If the Drude model approach is used,
this difference is  approximately 2.6\,pN (i.e.,
it is almost 9 times larger than the experimental
absolute error), thus, upsetting the agreement between
experiment and theory.

\subsection{Dynamic experiment using a microelectromechanical
torsional oscillator}

In the second, dynamic, experiment of Ref.~\cite{31} the
equivalent
Casimir pressure between two parallel plates (one coated by Au
and another by Cu) was determined dynamically in the
separation range from 260 to 1200\,nm with an absolute error
of about 0.6\,mPa. In the comparison
between the experimental data and different theoretical
approaches, a number of possible corrections to the theoretical
Casimir pressures were taken into account. These include those
due to surface roughness, the use of the proximity force
theorem, the finite sizes of the plates, sample-dependent
variations of the tabulated optical data for the complex
index of refraction, and contribution of nonlocal effects.
The differences between the theoretical and experimental Casimir
pressures, 
$\left[P^{\rm th}(z)-P^{\rm exp}(z)\right]$, deviate from
zero in the range from --0.8 to 0.5\,mPa if the
traditional theoretical approaches to the thermal Casimir
force are used \cite{31}. This demonstrates that the
traditional approaches are consistent with the experimental
data. If, however, the theoretical Casimir pressures
${\tilde{P}}^{\rm th}(z)$ are computed using the Drude
model approach, the differences 
$\left[{\tilde{P}}^{\rm th}(z)-P^{\rm exp}(z)\right]$ 
deviate significantly
from zero within the separation region from 260 to 700\,nm.
At the shortest separation this deviation reaches 5.5\,mPa
(i.e., it is more than 9 times larger than the experimental
absolute error). In Ref.~\cite{31} the conclusion was drawn
that the Drude model approach is ruled out experimentally.

\subsection{Improved dynamic experiment using a 
microelectromechanical torsional oscillator}

In Refs.~\cite{32,33} the results of a new, improved, dynamic
experiment were reported on the determination of the Casimir
pressure between the two plates both coated by Au in the
separation region from 160 to 700\,nm. The main improvements
were a significant suppression of the surface roughness,
a decrease by a factor 1.7 in the error in the
measurement of the absolute separations, and the use of
rigorous statistical procedures in data processing and in
the comparison of experiment and theory. Importantly,
the contribution of the surface roughness to the Casimir
pressure was reduced to less than 0.65\% even at the 
shortest separation. This is the first
experiment where the total relative error of the Casimir
pressure measurements found at 95\% confidence varies
between 0.54 and 0.59\% in a wide separation region from
170 to 300\,nm, and is as small as 2.5\% at $z=500\,$nm.

In addition to the above-mentioned corrections in the
theoretical Casimir pressure, the contribution of patch 
potentials and errors in the calculated pressures, arising
from the uncertainties in the experimental distances
\cite{34} were taken into account. Finally, the total
theoretical error, which was determined at 95\% confidence, 
decreases from 1.65\% at $z=160\,$nm to 0.9\% at
$z=750\,$nm. Using the total experimental and  
theoretical errors, determined independently, the 
absolute error of the pressure differences 
$\left[P^{\rm th}(z)-P^{\rm exp}(z)\right]$ was found at 95\%
confidence. It is practically the same in different
theoretical approaches. The solid lines in Fig.~1(a,b)
show the confidence interval versus
separation which is determined in terms of this error. 
In Fig.~1(a) the differences 
$\left[P^{\rm th}(z)-P^{\rm exp}(z)\right]$
for the traditional theories are plotted. In this case more 
than 95\% of all individual points (not only the mean values
of pressure differences at each separation as is demanded
by the theorems of mathematical statistics) belong to the
confidence interval. This means that our error analysis
is in fact too conservative and the errors are significantly
overestimated. From Fig.~1(a) it follows that the
traditional approaches to the thermal Casimir force are
consistent with data. In Fig.~1(b) the theoretical
pressures ${\tilde{P}}^{\rm th}(z)$ are computed using
the Drude model approach. It is seen that this
approach is excluded experimentally within the separation
region from 170 to 700\,nm at 95\% confidence.
As is shown in Ref.~\cite{32}, in the separation region
from 300 to 500\,nm the Drude model approach is excluded by
experiment at even higher 99\% confidence.
%%%%%%%%%%%%%%%%%%%%%%%%%%%%%%%%%%%%%%%%%%%%%%%%%%%%%%%%%%%%%%
\begin{figure}[t]
\vspace*{-6.5cm}
\includegraphics{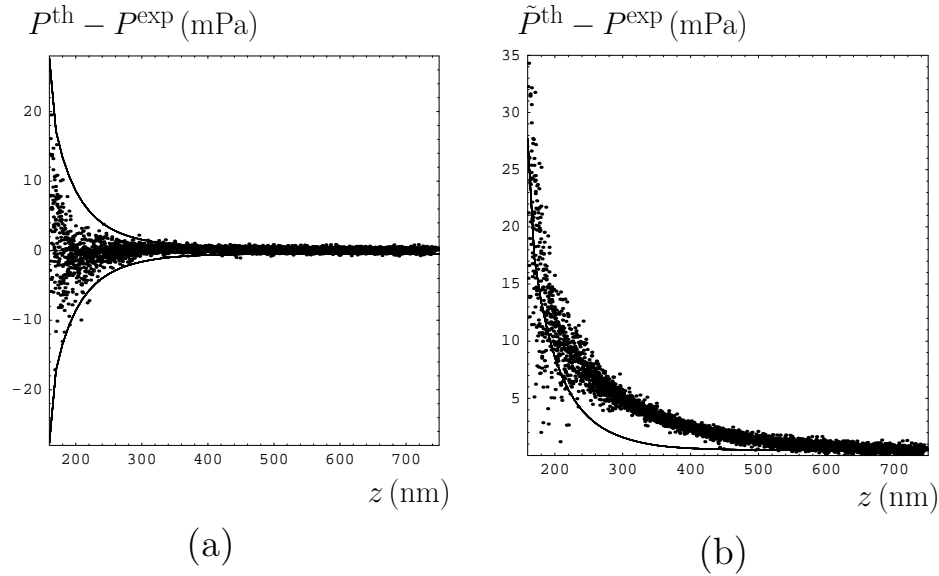}
\vspace*{-15.5cm}
\caption{
Differences between the theoretical Casimir pressures
calculated using one of the traditional approaches (a)
or the Drude model approach (b) and experimental
ones versus separation. The 95\% confidence interval lies
between the solid lines. 
}
\end{figure}
%%%%%%%%%%%%%%%%%%%%%%%%%%%%%%%%%

In the first version of preprint \cite{24}, it was claimed that
the computations of Ref.~\cite{32} in the framework of the
Drude model approach deviate somewhat from the original
computations of Ref.~\cite{32}. Later it was shown \cite{23}
that these deviations were caused by numerical
discrepancies in the data used in preprint \cite{24}, which
was recognized by the authors 
in the second version of this preprint.

Nevertheless, Ref.\cite{24} claims that ``the experimental
situation is still too indecisive to draw definite conclusions''
on the contradiction of the Drude model approach with experiment.
To justify this conclusion, Ref.~\cite{24} cites
Ref.~\cite{34} where the influence of the experimental
uncertainty in surface separations on the theoretical values
of the Casimir force was discussed. As was mentioned above, the
full contribution from this 
effect is included in the error analysis of
Ref.~\cite{32}, and thus cannot be used to cast doubts on the
accuracy of the obtained results. Ref.~\cite{24} 
also claims that ``the roughness of the surfaces is
much larger than the precision stated in the determination of
the separation''. This, however, cannot compromise the
precision achieved in Ref.~\cite{32} because the separations
are measured between zero levels of surface roughness
independently of the value of roughness amplitude. Furthermore,
the contribution of roughness in the experiment of Ref.~\cite{32}
is so small, that the validity of all the conclusions is
preserved even if one were to neglect the roughness in the
theoretical analysis. As was demonstrated in Ref.~\cite{23},
no unaccounted fixed systematic error in surface separation
is capable of bringing the Drude model approach into
agreement with data within the whole range of measurements.
One more claim of Ref.~\cite{24} is that ``accurate determination
of a small difference between experimental values at room
temperature and purely theoretical values at $T=0$ gives rise
to further difficulties''. This claim is evidently misleading
because all theoretical values of the Casimir pressure
in Ref.~\cite{32} were computed at $T=300\,$K, i.e., at the
experimental temperature. To summarize, all of the arguments of
Refs.~\cite{12,24} against the precision and accuracy of
the experiment of Ref.~\cite{32} break down, and the
conclusion that there is a contradiction between the Drude model
approach and experiment seems unavoidable.

\section{Spatial dispersion and the Casimir force}

Recent papers \cite{35,36} claim that taking account of the
spatial dispersion is crucial for the problem of the
thermal Casimir force between real metals. They use the 
usual (continuity) boundary conditions (\ref{eq3}) and
the usual Lifshitz formula (\ref{eq1}) with the reflection
coefficients expressed it terms of the spatially
nonlocal dielectric permittivities 
$\varepsilon_i(\omega,k_{\bot})$ or impedances
$Z_{\|,\bot}(\omega,k_{\bot})$ (the latter are presented
in terms of nonlocal dielectric permittivities by means of
some integral relations). Refs.~\cite{35,36} arrive at
the conclusion that the contribution from  
$r_{\bot}(0,k_{\bot})$ is negligibly small,
leading to approximately
the same results for the thermal Casimir force
as in the Drude model approach. In contrast to the Drude model
approach, entropy was shown to be equal to zero at $T=0$ even
for perfect crystal lattices \cite{36}.

As was already demonstrated in Sec.~4, the predictions of 
the Drude model approach are in contradiction with experiment.
There are also serious theoretical objections against the
formalisms used in Refs.~\cite{35,36}. It has been known
that in the presence of spatial dispersion the continuity
boundary conditions must be generalized for \cite{37,38}
\begin{equation}
\fl
E_{1t}=E_{2t}, \quad
B_{1n}=B_{2n}, \quad
D_{2n}-D_{1n}=4\pi\sigma,\quad
[\mbox{\boldmath{$n$}}\times({\mbox{\boldmath{$B$}}}_2-
{\mbox{\boldmath{$B$}}}_1]=\frac{4\pi}{c}\mbox{\boldmath{$i$}},
\label{eq19}
\end{equation}
\noindent
where
\begin{equation}
\sigma=\frac{1}{4\pi}\int_{1}^{2}
\mbox{div}[\mbox{\boldmath{$n$}}\times
[\mbox{\boldmath{$D$}}\times\mbox{\boldmath{$n$}}]]dl,
\quad
\mbox{\boldmath{$i$}}=\frac{1}{4\pi}\int_{1}^{2}
\frac{\partial \mbox{\boldmath{$D$}}}{\partial t}dl
\label{eq20}
\end{equation}
\noindent
are the induced charge and current surface densities (assuming 
there are no external charges and currents). If spatial
dispersion is absent, then $\sigma=0$, {\boldmath{$i$}}$=0$,
and the usual continuity boundary conditions (\ref{eq3}) are
recovered \cite{37,38}. Then the use of boundary conditions
(\ref{eq3}) in the presence of spatial dispersion is unjustified.

Refs.~\cite{35,36} disregard the fact that spatially
nonlocal dielectric permittivities $\varepsilon_i(\omega,k_{\bot})$
can be rigorously introduced only for an unbounded medium which
is uniform in space \cite{37,38}. In the theory of the anomalous
skin effect, such permittivities are sometimes used in the
presence of boundaries (see, e.g., Ref.~\cite{40a}). This is
an approximation applicable to the investigation of some bulk
effects, when physical phenomena caused
by a layer adjacent to the boundary
surface are neglected. (There is another approach 
to the theory of the
anomalous skin effect \cite{40b} which only deals with the 
Leontovich impedance.) In any case, it is unlikely that the
approximate phenomenological approach using nonlocal
$\varepsilon_i(\omega,k_{\bot})$ in the presence of boundaries 
would be applicable in the theory of Casimir force between
metal surfaces, where the boundary effects are of prime
importance.

It has also been known \cite{39} that with inclusion of 
spatial dispersion the free energy of a fluctuating field
${\tilde{\cal F}}(z,T)$ takes a more general form than
is given by the Lifshitz formula, i.e.,
\begin{equation}
{\tilde{\cal F}}(z,T)={\cal F}(z,T)+\Delta{\cal F}(z,T),
\label{eq21}
\end{equation}
\noindent
where ${\cal F}(z,T)$ is the Lifshitz contribution presented
in Eq.~(\ref{eq1}), and the general expression for
$\Delta{\cal F}(z,T)$ in terms of the thermal Green's functions
of the electromagnetic field can be found in Ref.~\cite{39}.
The use of the Lifshitz free energy $ {\cal F}(z,T)$ for 
systems with spatial dispersion (as was done, e.g., in 
Ref.~\cite{40} where $\varepsilon_i(\omega,k_{\bot})$ of the
hydrodynamic model 
were substituted into ${\cal F}$) is not correct \cite{39}.
As was emphasized in Ref.~\cite{39}, ``For most of condensed
matter bodies this is inadmissible''. 
It seems, however, that Refs.~\cite{35,36} repeat this misuse
of the Lifshitz formula.
At the same time, it is
amply evident that at $T=300\,$K in the region of infrared
optics (i.e., at the experimental separations) or in the
region of the normal skin effect ($z>4-5\,\mu$m) spatial 
dispersion does not play any role and can be neglected.

\section{Conclusions}
From the above the following conclusions can be formulated:

--- The thermal Casimir force between real metals presents 
serious problems, and there is presently no fundamental theory
which describes it;

--- The Drude model approach to the thermal Casimir force 
violates the Nernst heat theorem in the case of metallic
perfect crystal lattices with no impurities and is excluded
by experiment;

--- The thermal effect in the Casimir force has not yet been 
measured. The traditional approaches to its description
based on the plasma model and on the Leontovich impedance
are consistent with experiment. 
Computations using the Lifshitz formula at $T=0$ and tabulated
optical data are also consistent with experiment;

--- The measurement of the small thermal effect, as predicted
by the traditional approaches, would be of great interest and
could help to discriminate between them. In this regard the
proposed experiments \cite{30,41,42} are very promising;

--- The desired fundamental theory of the thermal Casimir
force between real materials should be presumably based on
a more sophisticated quantization procedure incorporating
spatial dispersion and inhomogeneity of space, and
may go beyond the scope of the Lifshitz theory.

%%%%%%%%%%%%%%%%%%%%%%%%%%%%%%%%%%%%%%%%%%%%%%%%%%%%%%%%

%%%%%%%%%%%%%%%%%%%%%%%%%%%%%%%%%%%%%%%%%%%%%%%%%%%%%%%
\end{document}